\documentstyle[aps,epsf]{revtex}
\begin{document}
\title{ FPU $\beta$ model: Boundary Jumps, Fourier's Law and
  Scaling} 
\author{Kenichiro Aoki\cite{ken-email}
  and Dimitri Kusnezov\cite{dimitri-email}
  }
\address{$^a$Dept. of Physics, Keio University, {\it
    4---1---1} Hiyoshi, Kouhoku--ku, Yokohama 223--8521, Japan\\
  $^b$Center for Theoretical Physics, Sloane Physics Lab, Yale
  University, New Haven, CT\ 06520-8120} \date{\today }
\maketitle
\begin{abstract}
We examine the interplay of surface and volume effects in
systems undergoing heat flow. In particular, we compute the
thermal conductivity in the FPU $\beta$ model as a function of
temperature and lattice size, and scaling arguments are used
to provide analytic guidance.  From this we show that boundary
temperature jumps can be quantitatively understood, and that
they play an important role in determining the dynamics of the
system, relating soliton dynamics, kinetic theory and Fourier
transport.
\end{abstract}
Transport in low dimensional systems has enjoyed renewed
interest due to new efforts to understand why some theories have
a bulk limit for transport coefficients and others do
not\cite{1dmodels,bhu,plb,km,llp}. The FPU $\beta$ model, in particular,
provides a classic example of a system with no bulk limit
\cite{km,llp}. Steady state transport is typically generated by
imposing boundary conditions on surfaces of the system of
interest, that provide the necessary behavior, such as shearing,
heat flow and so forth.  A physical and ubiquitous feature of these
boundary-driven non-equilibrium steady states are surface
discontinuities in certain observables\cite{physical-jumps}.  In
fluids that are being sheared, a slip velocity develops at the
boundary reflecting the fact that the fluid just inside the
system does not move with the same velocity as the walls. In
systems undergoing heat flow, as in the FPU model we study here,
a `surface resistance' or boundary jump in temperature develops.
We will show that these surface effects can be quantitatively
understood, and that the interplay of surface and volume effects can
be used to define dynamical regimes, characterized by different
scaling properties. To provide a more complete picture, we compute
both the temperature and size  dependence of the thermal
conductivity $\kappa$, as well as the speed of sound and other
kinetic theory quantities. From the behavior of $\kappa$, we will also
see that the presence of a constant thermal gradient does not
guarantee that Fourier's law, specifically the linear relation
between the heat flow $J$ and $\nabla T$, $J=-\kappa\nabla T$,
is valid.  (We refer to the relation $J=-\kappa\nabla T$ as  Fourier's
law, even in the absence of the bulk limit.) The role of
solitons can also be understood.

Heat flow is generated by applying hot ($T_2^0$) and cold
($T_1^0$) sources to the different boundaries of the
system. However, in general there will be a 
mismatch between the surface temperature $T^0_i$ and 
temperature just inside the system, $T_i$, denoted $\delta
T_i=|T_i-T^0_i|$\cite{physical-jumps}. Consequently, 
the {\sl input} temperature difference, $\Delta T=T_2^0-T_1^0$,
can be decomposed into the two surface jumps, $\delta T_1$, $\delta
T_2$, and the gradient contribution: 
$\Delta T=\int dx\,\nabla T+ \delta T_1 +\delta T_2$. 
The jumps $\delta T_i$ can be expressed as $\delta
T_i=\eta \nabla T|_i$, where $\nabla T_i$ is the internal
temperature gradient extrapolated to the surface $i$, and $\eta$
is on the order of the mean free path at that temperature $\lambda_i$
\cite{pk,pla}. Hence we write 
$\delta T_i = \alpha \lambda_i \left.\nabla T\right|_i$, where
$\alpha$ is a constant corresponding to the efficacy of the
boundary thermostats as they couple to the system. In this work,
it will turn out 
that the boundary jumps become important in the near equilibrium
regime, so that they are symmetric, $\delta T_1\simeq\delta T_2$, and the
gradient is constant, $\nabla T_i=\nabla T$. Thus we are led to
the general near--equilibrium relationship, using the length $L$
between the thermostats,
\begin{equation}
  \label{eq-dt}
 \left[ 2\alpha\lambda + L\right] \nabla T = \Delta T.
\end{equation}
Seemingly simple, this relation can be used to derive
many properties of the FPU model, and others in general.

In this letter we examine the FPU $\beta$ Hamiltonian,
${\tilde H}  = \sum_{k=1}^L \left[
  \frac{\tilde p_k^2}{2m}+\frac{1}{2}m\omega^2
 (\tilde q_{k+1}-\tilde q_k)^2 + \frac{\beta}{4} (\tilde
  q_{k+1}-\tilde q_k)^4\right].
$
Under the rescaling $\tilde p_k=p'_k\omega^2\sqrt{m^3}$, $\tilde
q_k= q'_k\omega\sqrt{m}$, we obtain the conventional
form of the FPU~$\beta$ model,
\begin{equation}\label{fpu-usual}
 H_\beta = \frac{1}{2}\sum_{k=1}^L \left[ p_k^{\prime 2}
  + (q'_{k+1}- q'_k)^2
  + \frac{\beta}{2} (q'_{k+1}- q'_k)^4\right],
\end{equation}
where $H_\beta=\tilde H/(m^2\omega^4)$. It is important to note
that the temperature and the coupling are {\it not} independent
parameters, but changing the temperature is equivalent to
changing the coupling $\beta$.  This can be seen by the further
rescaling 
$p'_k=p_k/\sqrt\beta $, $q'_k=q_k/\sqrt\beta$, which leads to a
unique, dimensionless, Hamiltonian $H\equiv H_{\beta=1}=\beta
H_\beta$.  Since we have rescaled the momenta $p_k^2=\beta
p_k'^2$, the temperatures in the two formulations $H$ and
$H_\beta$ are related by
$T=\beta T'$. So the most general physics can be obtained by
using $H_\beta$ around a fixed temperature (for instance $T'=1$)
but with different couplings $\beta$, or by using $H$ at various
temperatures $T$. In this study we use $H$ and probe
temperatures from $10^{-4}$ to $10^4$, where 
$T_k=\langle p_k^2\rangle$ (ideal gas thermometer),  using
$L$ sites and either free
($q_0=q_1$, $q_L=q_{L+1}$) or fixed ($q_1=q_L=0$) boundary
conditions. Perhaps surprisingly, these conditions have an
effect on the measured transport properties.

Heat flow is generated when thermostats are coupled to the
boundaries of the system. The system evolves according to
Hamilton's equations of motion
except at the boundaries, where we use Nos\'e-Hoover (NH)
thermostats\cite{nose-hoover} or demons\cite{thermostats} 
(see \cite{pla}, Eqs. (3)).
The NH thermostats are observed not to thermalize the boundaries
for large temperature gradients, a problem particularly acute
for small lattices. The demons, on the other hand,  do not
suffer from this problem and hence provide a resolution to the
problem recently studied in \cite{fillipov}.
We have  considered a range of coupling strengths as well as
thermostatting one or more sites on each edge. Irrespective of
the form of the boundary thermostats, we note that their
introduction breaks global momentum conservation,
$\dot P_{tot}\not=0$, while it is
conserved on average; $\langle \dot P_{tot}\rangle =0$.
The equations of motion, on the other hand, are still invariant
with respect to the translations $q_k\rightarrow q_k+c$ for
free or periodic boundary conditions, where
$c$ is an arbitrary constant. This 
disparity arises from the fact the the equations of motion for
the whole system are no longer Hamiltonian. 

\begin{figure}
\begin{center}
    \leavevmode
    \epsfxsize=8cm\epsfbox{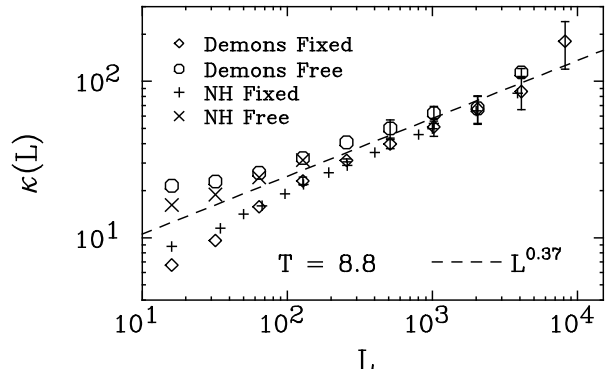}
  \caption{Size dependence of the thermal conductivity of the FPU model
    for free or fixed boundary conditions with different
    thermostats. The Nos\'e-Hoover thermostats do not thermalize
    the boundaries for small $L$ and hence their results deviate
    from those of the demons. $L^{0.37}$ behavior will be
    discussed along with Fig.~\ref{fig2}.}
\label{fig1}
\end{center}
\end{figure}
The definition of heat current $J$  depends on the form of
the discrete gradient in the Hamiltonian and continuity
equations. For instance, 
we may choose $h_k=p_k^2/2 +U(q_{k+1}-q_k)$ (where $H=\sum_k
h_k$, $U(x)=x^2/2+x^4/4$) and $\dot h_k+J_{k+1}-J_k=0$, leading
to the consistent form $J_k=-p_k U'(q_k-q_{k-1})$. Regardless of
the form chosen, the measurement of $J$ is the same. 
We then use  Fourier's law,
$J=-\kappa(T,L)\nabla T$, to obtain $\kappa$.
In Fig.~\ref{fig1} (NH-fixed), we reproduce the analysis of
\cite{llp},  plotting $\kappa$ using the boundary temperatures
$(T_1^0,T_2^0)=(2.4,15.2)$ to 
obtain $\kappa(8.8,L)$ and compare this to the results obtained
using demons and also the free boundary condition results using
the same $(T_1^0,T_2^0)$.
This figure raises several significant points. 
First, the NH results do not thermalize 
the boundaries for smaller $L$, which accounts for the disagreement with
the demon results.  There is also a
clear difference between the 
fixed or free boundaries.  Free boundary conditions are found to
result in the diffusion of the {\sl entire} system in a Brownian
manner: $\langle q^2\rangle\sim t$, an essential
difference. The 
differences in $\kappa$ disappear as we increase the system
size since the temperature gradients become small and we reach the
near equilibrium regime. 
For small $L$,  while $\nabla T\sim$constant, the
system does not obey Fourier's law. We return to this in
Fig.~\ref{fig:fourier} 
below. Consequently, referring to these points as $\kappa$ is not
entirely correct. 

\begin{figure}
\begin{center}
    \leavevmode
    \epsfxsize=7.5cm\epsfbox{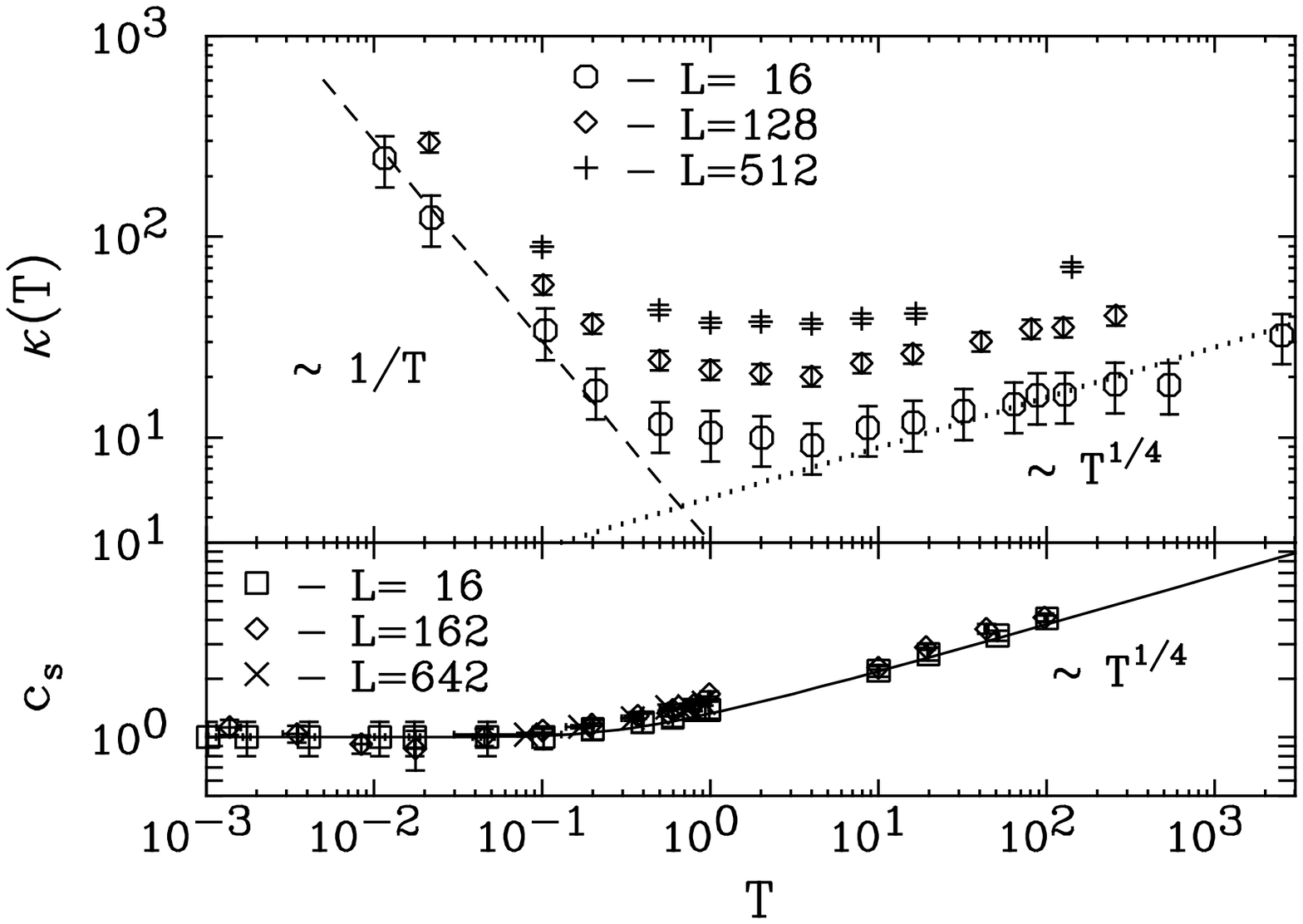}
    \caption{(Top) $T$ and $L$ dependence of $\kappa$.
    One can 
    see two regimes. For large $T$, there is an approximate
    $T^{1/4}$ behavior which can be understood from scaling
    arguments. For low 
    $T$, a $1/T$ dependence which arises due to the weak
    coupling limit. The region where $T\sim 10$ has $\kappa$
    largely independent of $T$, so that Fourier's law can be
    strongly violated while the temperature profiles remain
    linear. This region was studied in [5] and Fig. 1. 
    (Bottom) Speed of sound compared to expectations
    from soliton kinetic energy. }
\label{fig2}
\end{center}
\end{figure}
In Fig.~\ref{fig2}~(top), we plot the $T$ dependence of $\kappa$
over 6~orders of magnitude for selected lattice sizes~$L$ with
the fixed boundary conditions.  
To obtain these results, we first verified that Fourier's law
$J=-\kappa(T,L)\nabla T$ is satisfied, by plotting $J$ versus
$\nabla T$ for various gradients. Only then, can we extract the
conductivity $\kappa$ correctly. 
The dotted and dashed lines represent the behavior we expect from
the following arguments:
At high $T$, the quartic term for $q_k$ in the
Hamiltonian~$H$ dominates over the quadratic term in
$q_k$, as can be seen from the virial theorem, for instance.
Therefore, in this asymptotic region, the Hamiltonian has the
scaling property, $H/T=H'/T'$ under the rescalings
$p_k=(T/T')^{1/2}p'_k$, $q_k=(T/T')^{1/4}q_k'$. In other words,
the physics at different temperatures is related by simple
kinematical rescalings.  Applying
this rescaling to the thermal conductivity, we derive $\kappa= -
\langle J\rangle/\nabla T=(T/T')^{1/4}\kappa'$, which well
describes the behavior $\kappa\sim T^{1/4}\ (T\gg1)$ we have
obtained.

The low temperature limit can be viewed as the weak coupling
limit ($\beta\ll 1$) of $H_\beta$. Near the harmonic limit, we expect
(and verify) the speed of sound $c_s$ and heat capacity $C_V$ to be
unity. Hence kinetic theory suggests that $\kappa\sim\lambda$,
where $\lambda$ is the mean free path. The mean free path in
this limit is on order of $\lambda\sim 1/\beta$\cite{ashcroft},
so that $\kappa\sim 1/T$.  We also find that the size dependence of
$\kappa$ has a power law behavior that is independent of the
temperature, which is consistent with the previous results
\cite{lepri-physicad,km}. Consequently, we expect and find
\begin{eqnarray}\label{eq-kap}
\kappa \simeq\cases{  1.2 L^\delta T^{-1}&   $(T\alt 0.1)$\cr
    2 L^\delta T^{1/4}&$ (T\agt 50)$\cr},\qquad
  \delta=0.37(3).
\end{eqnarray}
This provides a much more global picture of heat
transport in the FPU model than previously known.

The excitations in the FPU model are known to be
solitonic\cite{bhu,solitons}. The `sound speed' $c_s$ can be
measured by examining how 
an initial perturbation in the center of the system
propagates, as a function of lattice size $L$ and temperature
$T$. In Fig.~\ref{fig2}~(bottom) we plot this together with the
velocity behavior suggested from solitons, namely, $2\, T=c_s^3
\sqrt{c_s^2-1}$\cite{solitons}.  The agreement is quite good
aside from a weak $L$ dependence which is not described by this approach.

\begin{figure}
\begin{center}
    \leavevmode
  \epsfxsize=7cm\epsfbox{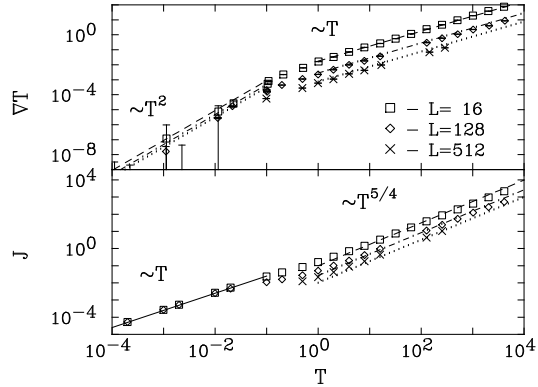}
  \caption{Behavior of $\nabla T$ and $J$ as a function of $T$ compared
    to expected behaviors. Solitons propagate freely below
    $T_c\sim L^{1-\delta}$.}
  \label{fig:JT}
\end{center}
\end{figure}
Consider now some consequences of Eq.~(\ref{eq-dt}).  Kinetic theory 
indicates that $\kappa=\lambda c_s C_v$. Since $C_v\sim1$ 
for all temperatures we measured, we have $\lambda\sim
\kappa/c_s$. Hence, at low $T$, $\lambda\sim 1/T$, and the
boundary jumps eventually dominate the thermal profile. The
transition to this regime occurs when
$L\sim 2\alpha\lambda = 2\alpha\kappa$ (we show below that
$\alpha\sim 1$), or using
Eq.~(\ref{eq-kap}), $T_c\sim 2 L^{1-\delta}$. 
To delineate this behavior, we study the system at
various central temperatures, $T$, with  fixed relative
differences $\Delta T/T\equiv r=0.4$ in Fig.~\ref{fig:JT}. 
For $T<T_c$, boundary jumps dominate Eq.~(\ref{eq-dt}), so 
that $J=\Delta T/(2\alpha)= (r/2\alpha)T\sim T/4$. Notice here
that in this regime, $J$ {\sl should be independent of} $L$. This is indeed
observed in Fig.~\ref{fig:JT}~(bottom). The transition
also is characterized by $\lambda \gtrsim L$, which means that
excitations (solitons) now move freely across the system
without interaction. This indicates the onset of the harmonic
limit which is characterized by a flat temperature profile. We
can use Eq.~(\ref{eq-dt}) to compute how fast we approach the
harmonic limit: 
$\nabla T= (r/2\alpha \kappa) T = 0.2 T^2 L^{-\delta}$. Hence the
system approaches the harmonic limit like $T^2$. We have
explicitly measured this behavior for several $L$ and confirm
it. 

A frequently used argument surmises that for fixed boundary
temperatures, the behavior of $\kappa(T,L)$ can be obtained from
the behavior of $J$ by assuming that $\nabla T$ scales like
$1/L$ (see for example \cite{tsironis,llp}). Our results
demonstrate that such an argument in general is not applicable
due to the presence of boundary jumps which cannot be ignored.
It should be emphasized that the  importance of jumps is 
unrelated to the absence of a bulk limit\cite{pla}.

At higher temperatures, boundary jumps compete with the heat
flow inside. Here $c_s$ and $\kappa$ both behave at $T^{1/4}$,
so that $\lambda$ is independent of $T$. Consequently, $\nabla
T=r/(2\alpha\lambda + L) T = [0.4/(L+3L^{\delta})] T$ and
$J=-\kappa\nabla T= [0.8/(L^{1-\delta}+3)] T^{5/4}$. In
Fig.~\ref{fig:JT},
we show both $\nabla T$ and $J$ as a function of $T$ and see
that the predicted behavior is consistent with measurements. 
The agreement includes not only the power law, but the
prefactors as well, both for low and  high $T$.
 
\begin{figure}
\begin{center}
    \leavevmode
  \epsfxsize=8cm\epsfbox{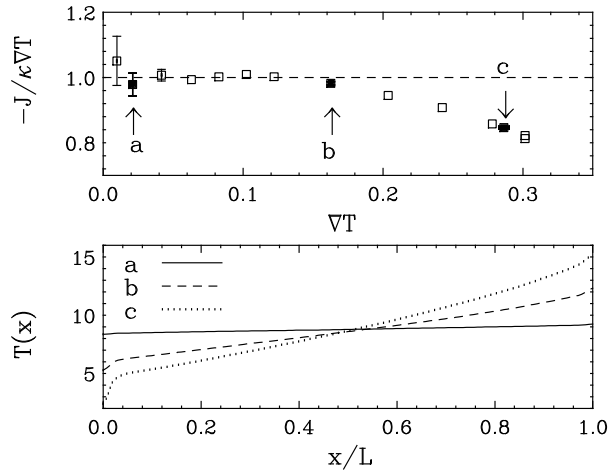}
  \caption{(Top) Violation of Fourier's Law, $-J/\kappa\nabla T$,
    as a function of $\nabla T$, showing departure from
    Fourier's law (dashes) at larger gradients. (Bottom)
    Temperature profiles  for the three points indicated in the
    top figure. One can see that the profile has a
    fairly constant  gradient in spite of the 20\% violation of
    Fourier's law in (c). }
  \label{fig:fourier}
\end{center}
\end{figure}
Recently we have derived a formula for the temperature profile
$T(x)$ in models where $\kappa(T)$ displays power law behavior,
$\kappa\simeq const./T^\gamma$, in the temperature range of
interest\cite{pla,plb}:
$  T(x) = T_1[1-(1-({T_2}/{T_1})^{1-\gamma})
    {{x}/{L}}]^{{{1}/{(1-\gamma)}}}.$
In the present study, in looking at
temperatures near T=8.8 in Fig. 2, $\kappa$ is nearly {\it temperature
  independent} over more than a decade in temperature. Hence we
expect $T(x)= T_1+(T_2-T_1)x/L + {\cal O}(\gamma)$ to develop
almost no curvature for very large
gradients, and hence the shape of $\nabla T(x)$ is {\it not} 
a good measure of the
validity of Fourier's law. In Fig.~\ref{fig:fourier} 
we plot $-J/\kappa\nabla T$ as a function of the measured  $\nabla T$. The
dashed line represent the expected result when Fourier's law
holds. One can see that as the gradient increases, noticeable
systematic violations of Fourier's law develop. For the selected
points, we plot the temperature profiles in
Fig.~\ref{fig:fourier} (bottom), demonstrating 
the nearly linear behavior. Similar behavior is also seen at higher 
temperatures, since $\kappa\sim T^{1/4}$ is a weak power law. 
Hence in contrast
the the $\phi^4$ model, where noticeable curvature in $T(x)$
develops when Fourier's law is broken\cite{plb}, the FPU model
demonstrates that {\it nearly constant temperature gradients do not
  imply Fourier heat flow.} It should noted that one will
underestimate the thermal conductivity if the Fourier's law
is applied in this regime to obtain the conductivity.

\begin{figure}
\begin{center}
    \leavevmode
  \epsfxsize=8cm\epsfbox{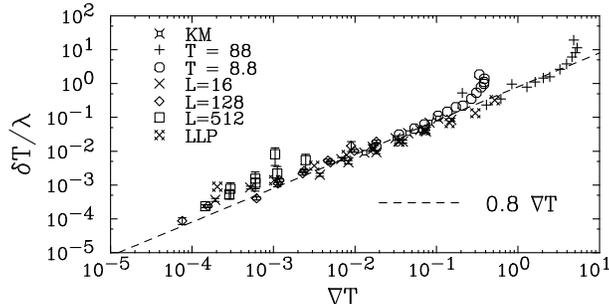}
  \caption{Quantitative study of boundary jumps $\delta T$ in
    our study, normalized by the mean free path $\lambda$, allow
    the determination of $\alpha$.
    The results from previous stochastic
    (KM) and dynamic (LLP) studies have been rescaled and are
    also shown. One can see that over an extremely large range
    in jumps and gradients, a simple scaling behavior exists.}
\label{fig:jumps}
\end{center}
\end{figure}
In order to compute the coefficient
$\alpha$, we must analyze the behavior of the boundary jumps
$\delta T/\lambda$ as a function of $\nabla T$. In
Fig.~\ref{fig:jumps},  we
combine results from various lattice sizes and temperatures, as
well as from stochastic boundary conditions (see for instance
Figs. 1 in ref. \cite{lepri-physicad} and 2-4 in
ref. \cite{km}).  
In Fig.~\ref{fig:jumps}, we plot $\delta T/\lambda$ and
see that this behaves as $\nabla T$. In addition we show the
behavior extracted from stochastic boundary conditions as well
as from previous calculations in the FPU, which are quite
consistent with our results. We then extract the 
parameter $\alpha=0.8(1)$ for the demons.  
A necessary consequence is that the
temperature profiles cannot be shape invariant in the way
discussed in \cite{llp}. 
While the same scaling is found for different thermostats, the
coefficient $\alpha$ can be different. For NH demons (LLP) used in
\cite{llp,lepri-physicad}, $\alpha_{NH}=2.0(3)$ and for
stochastic boundary conditions (KM) in \cite{km}, it is much
larger, $\alpha_{S}=40(6)$, as evidenced in the large boundary
jumps seen in \cite{km}. 

We have seen that the interplay of surface and volume effects in
Eq. (1) can be used to understand transport properties.
Since the conductivity in this model has no bulk limit, the
transport depends 
not only on the method used to thermalize the boundaries, but
also on the boundary conditions themselves. Nevertheless, one
can understand the temperature dependence at low and high
temperatures. 
We have been able to map out $\kappa(T,L)$ and
provide an analytical understanding in the asymptotic regimes.
The FPU model also provides an example of a theory where
Fourier's law breaks down, but the temperature profile retains a
nearly constant gradient. This can be understood in terms of the
weak temperature dependence of the conductivity. The boundary
jumps also display a very systematic behavior over many decades
in $\nabla T$, independent of the types of thermostats (dynamics
or stochastic) or of the lattice size. This dependence on
$\nabla T$ also shows that the temperature profiles cannot be
shape invariant unless the boundary jumps are explicitly taken
into account.

We acknowledge support 
from grants at Keio University and DOE grant DE-FG02-91ER40608.

\end{document}